\documentclass{optica-article}

\journal{oe}

\articletype{Research Article}

\begin{document}

\title{Polarization Imaging of Back-Scattered Terahertz Speckle Fields}

\author{Kuangyi Xu,\authormark{1} Zachery B. Harris,\authormark{1} and M. Hassan Arbab\authormark{1,*}}

\address{\authormark{1}Department of Biomedical Engineering, State University of New York at Stony Brook, Stony Brook, New York 11794, USA}

\email{\authormark{*}hassan.arbab@stonybrook.edu} 

\begin{abstract}
Speckle patterns observed in coherent optical imaging reflect important characteristic information of the scattering object. To capture speckle patterns, angular resolved or oblique illumination geometries are usually employed in combination with Rayleigh statistical models. We present a portable and handheld 2-channel polarization-sensitive imaging instrument to directly resolve terahertz (THz) speckle fields in a collocated telecentric back-scattering geometry. The polarization state of the THz light is measured using two orthogonal photoconductive antennas and can be presented in the form of the Stokes vectors of the THz beam upon interaction with the sample. We report on the validation of the method in surface scattering from gold-coated sandpapers, demonstrating a strong dependence of the polarization state on the surface roughness and the frequency of the broadband THz illumination. We also demonstrate non-Rayleigh first-order and second-order statistical parameters, such as degree of polarization uniformity (DOPU) and phase difference, for quantifying the randomness of polarization. This technique provides a fast method for broadband THz polarimetric measurement in the field and has the potential for detecting light depolarization in applications ranging from biomedical imaging to non-destructive testing.
\end{abstract}

\section{Introduction}
Speckle patterns are usually formed when a coherent light reflects or transmits through a rough surface or a turbid medium. Their unique spatial feature is described by a random granular structure. Speckle patterns can be considered a hurdle that degrades the image quality, but they also carry the characteristics information regarding the illuminated scattering object and have enabled innovative breakthroughs in many imaging applications. In recent years, speckle patterns with unique and tailored statistical properties \cite{Bender:18}, often described by their probability density function (PDF), have found promising applications in microscopy \cite{Mudry:12, Lim:08}, super-resolution imaging \cite{Redding:13,Katz:14,Lai:15}, ghost imaging \cite{Pelliccia:16}, Speckle-Tracking Echocardiography \cite{Geyer:10,Li:20}, and diffuse biomedical spectroscopy and imaging \cite{Mariampillai:08,Roustit:10,Aizawa:11}.

Coherent terahertz (THz) spectroscopy has likewise enjoyed a wide range of promised biomedical \cite{Goretti:17,Sung:18,Wang:21,Chen:21} and non-destructive testing applications \cite{DONG:15,Zhong:19}, which can give rise to scattering and speckle phenomena at different wavelength scales. Some of the first investigations of this phenomenon in the THz regime were studied by analyzing the statistics of the amplitude and phase of the broadband diffuse scattering \cite{Jian:03,Pearce:03}. The frequency-dependent scattering loss of THz pulses has been reported for different sample morphology, including the surface scattering due to roughness \cite{Arbab:10,Khani:20,Khani:21-2} and the volume scattering in porous \cite{Zhao:02,Xing:19} or granular \cite{Zurk:07,Khani:21} mediums. There has been significant electromagnetic modeling and signal processing efforts in relating the THz spectra with limited parameters describing the samples through turbid media  \cite{Bandyopadhyay:07,Dean:08,Mayank:12,Osman:22}, which usually require approximations of the dielectric properties in order to distinguish between scattering and feature-less absorption losses. The alternative approach is to investigate the unique phenomena of scattering, such as the optical diffusion measured via angular-resolved detection \cite{Osman:19}, the change in polarization states characterized by the Mueller matrix of samples \cite{Notake:19,Huang:20}. However, these methods are not commonly used in THz spectroscopy systems due to cumbersome alignment and time-consuming nature.

THz speckle patterns with broadband spectroscopic or polarimetric signatures have not been investigated due to technological challenges. For example, monochromatic THz speckle images with limited bandwidth were captured using a free-electron laser operating at 2.3 THz \cite{Vinokurov:09}, and also reported in the imaging by radar systems working at around 600 GHz \cite{Petkie:12,Grossman:14}. Speckles constructed by a broadband THz illumination were captured through the combination of the electro-optic sampling with a CMOS camera \cite{Leibov:21}. However, spectroscopic signatures of the speckle were not investigated. In order to characterize the THz speckle  occurred in practical applications, it is necessary to extend the single-point terahertz time-domain spectroscopy (THz-TDS) into a high-speed and portable spectral imaging technique. For instance, to capture \textit{in vivo} THz speckle patterns in biophotonics applications, the instrument should be robust and rapid such that it does not introduce additional grainy features due to the motion artifacts of the subject or the user. Further, the polarization change information of the THz beam can be used to discern biological information with higher sensitivity through light scattering mechanisms \cite{Chen-ADPR:21}. We have developed single-point THz time-domain polarimetry (THz-TDP) techniques with high measurement accuracy over a broad spectral bandwidth \cite{xu:2020, Xu:2022}, however they lacked beam steering or image formation capability. Until recently, broadband THz-TDS spectral or polarimetric imaging techniques that can achieve large field of view (FOV) and fine spatial resolution necessary for studying speckle patterns were lacking.

In this paper, we present a polarization-sensitive and fast imaging method to directly measure the spatial distribution of the THz speckle fields. Previously, we have developed a PHASR (Portable HAndheld Spectral Reflection) Scanner, which adopted a telecentric beam steering strategy, to image the full 40 × 27 mm FOV in both Asynchronous Optical Sampling (ASOPS) and Electronically Controlled Optical Sampling (ECOPS) modes \cite{Zachery:20,Zachery:22}. By incorporating two Photoconductive Antenna (PCA) detectors and a polarizing beam-splitter to simultaneously record the two orthogonal polarization directions of the THz field, we have upgraded our PHASR Scanner to capture polarization-sensitive images of the target without sacrificing the scanning speed. We investigated the polarization speckle patterns formed by the Stokes vectors calculated from the THz-TDS images of gold-coated sandpapers. We observed that the polarization of the back-scattered THz fields became either partially (under-developed speckles) or fully randomized according to the degree of roughness and the wavelength of the THz illumination. 
The statistical properties of the back-scattered THz speckle fields are in agreement with the non-Rayleigh models of partially-developed speckles, which are valid when the sample roughness is smaller than the wavelength of illumination. Our results show that the normal-incidence THz polarimetric reflection imaging modality can be used for characterizing the depolarization caused by rough surface, which provides the potential for its applications in other turbid media and highly scattering samples. 

\section{Methods}

\begin{figure}[b!]
\centering\includegraphics{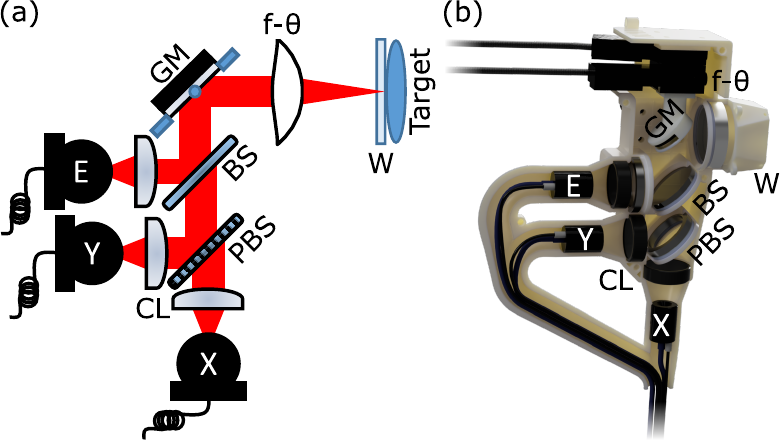}
\caption{(a) Schematic and (b) photograph of of the optical components inside the two-channel PHASR Scanner housing. }
\label{schematic}
\end{figure}

The THz-TDS measurements are obtained using an upgraded version of our PHASR (Portable HAndheld Spectral Reflection) Scanner, described in detail previously \cite{Zachery:22}. The upgraded PHASR Scanner, shown in Fig.  \ref{schematic}, includes a polarizing beam splitter and two PCA detectors oriented along orthogonal directions. The PHASR Scanner incorporates the TERA ASOPS (Asynchronous OPtical Sampling) dual-fiber-laser THz spectrometer (Menlo Systems, Inc., Newton, NJ, USA) into a handheld, collocated, telecentric imaging system. A THz beam generated by the photoconductive antenna (PCA) in the emitter (E) is collimated using a TPX lens (CL) with 50 mm focal length. The collimated beam is directed towards a gimballed mirror (GM) using a high-resistivity silicon beam splitter (BS). The gimballed stage is a two-axis motorized system composed of a goniometer and a rotational stage. It raster scans the collimated beam over the aperture of a custom-made telecentric $f\textrm{-}\theta$ lens \cite{Zachery:19}. Therefore, the focused beam is always normal incident onto the target and has a constant focal spot-size. A free-standing grid acts as a polarizing beam splitter (PBS) after the BS, which separates the back-scattered (or specularly reflected) radiation into the two orthogonal components denoted by X and Y. The signals from the two PCA detectors are converted in two transimpedance amplifiers, and then collected with two digital acquisition (DAQ) cards at synchronized time.

\begin{figure}[b!]
\centering\includegraphics{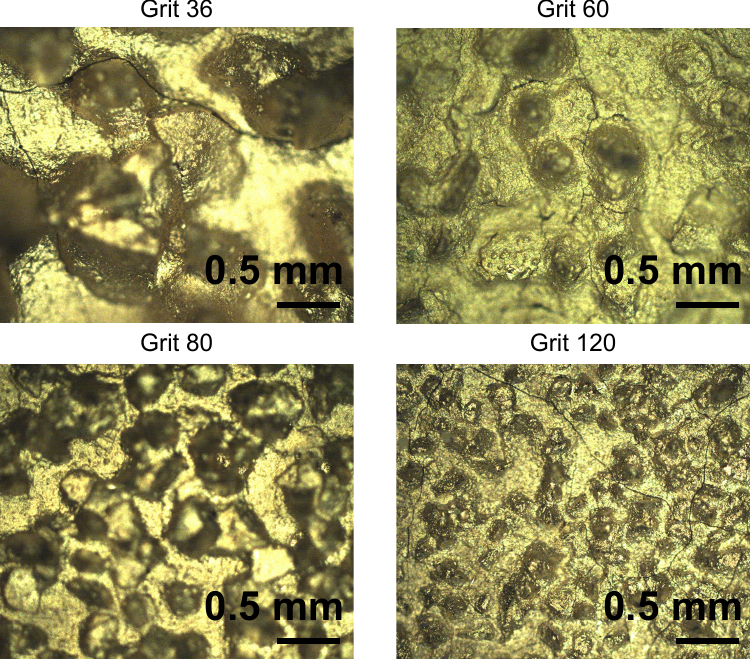}
\caption{Microscope images (5x) of gold-coated sandpapers of grit 36, 60, 80 and 120.}
\label{fig:sandpaper}
\end{figure}

To investigate the depolarization of THz waves due to scattering, we prepared targets with high reflectivity and different surface roughness. These targets are made of sandpaper of grits 36, 60, 80 and 120 (lower grit number corresponds to larger average particle size and thus rougher sandpaper \cite{sandpaper}), each coated with a 120-nm layer of gold by vacuum deposition. Figure \ref{fig:sandpaper} shows the microscope images of the gold-coated sandpapers with 5x magnification. These sandpapers are placed at the focal plane of the $f\textrm{-}\theta$ lens, scanning by 20×20 $mm^2$ FOV with the pixel size of 0.5×0.5 $mm^2$. Each pixel is recorded in 1 seconds by averaging 100 replicates in time-domain. Images of a flat mirror are taken under the same setting as references to establish the minimum polarimetric detection resolution of our instrument and the laser-induced speckle background. In the following measurements, the emitter PCA was oriented to 45 degrees to ensure sufficient THz power in both the X and Y detection channels.

Subsequently, the averaged THz-TDS waveforms for components X and Y are converted to complex frequency functions $\mathbf{A}_x$ and $\mathbf{A}_y$ via Fourier transform. These complex functions can be represented in terms of the real Stokes parameters by \cite{Goodman:20},
\begin{equation}
\centering
\begin{split}
I=\mathbf{A}^*_x\mathbf{A}_x+\mathbf{A}^*_y\mathbf{A}_y,\;Q=\mathbf{A}^*_x\mathbf{A}_x-\mathbf{A}^*_y\mathbf{A}_y,\\
U=\mathbf{A}^*_y\mathbf{A}_x+\mathbf{A}^*_x\mathbf{A}_y,\;V=i(\mathbf{A}^*_y\mathbf{A}_x-\mathbf{A}^*_x\mathbf{A}_y).
\label{eq:1}
\end{split}
\end{equation}
The Stokes parameters derived above has less degree of freedom than those obtained from intensity measurement since $I^2=Q^2+U^2+V^2$ is guaranteed by Eq. \ref{eq:1}, which would account for an apparent fully-polarized THz wave regardless of its actual states. However, it is well known that light can appear depolarized on average even if it is fully polarized locally \cite{Ellis:04}. The imaging capability of our instrument makes it possible to characterize the spatial variation of the Stokes parameters, which is closely relevant to the surface morphology of the sandpapers.

\begin{figure}[b!]
\centering\includegraphics{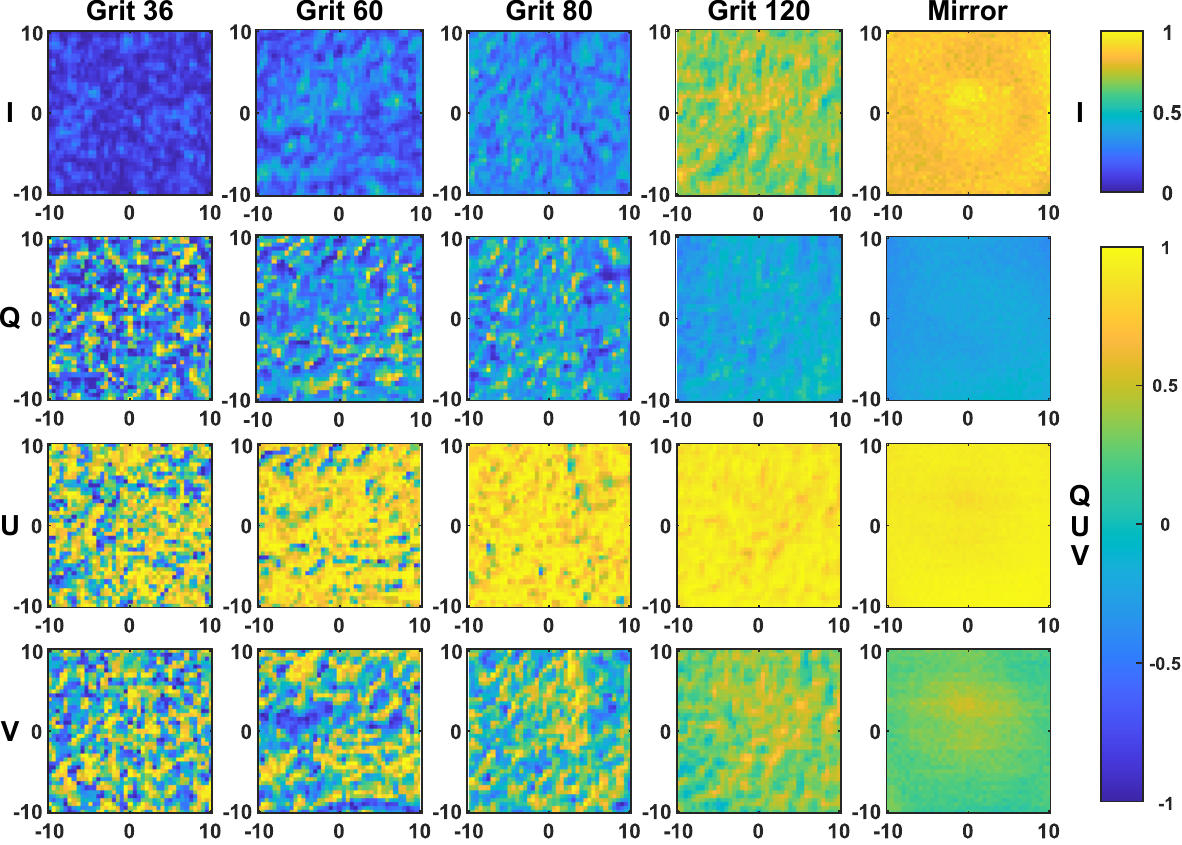}
\caption{The spatial variation of Stokes parameters $I$, $Q$, $U$ and $V$ for the speckles formed by different targets at 0.5 THz. $I$ is normalized by the maximum pixel value in the image of the mirror, as the reference, while the other Stokes parameters are normalized by $I$.}
\label{fig:IQUV}
\end{figure}

\section{Results and Discussions}
\subsection{Imaging of THz Speckle Fields with Stokes Parameters}
Figure \ref{fig:IQUV} presents the spatial variation of Stokes parameters in the THz images obtained by normal-incidence backscattered wave from the surface of five targets, including the gold-coated sandpapers of four different grit numbers and a flat mirror. The signals reflected by mirror are considered uniform in the contrasts of the Stokes parameters, despite some minor variations accounted by the intrinsic systematic errors of our instrument. As the surface roughness of the target increases, the images gradually become more inhomogeneous and contain granular structures. The granularity of parameter $I$ is a known feature of speckle noise, while the decrease of $I$ over increasing target roughness suggests the transition from specular reflection to back-scattering. Meanwhile, the spatial fluctuations of $Q$, $U$ and $V$ implies the random nature of the polarization states for the back-scattered THz fields. It can be observed that the THz speckle images using Stokes parameters are closely relevant to the surface structures of the targets. 

\begin{figure}[b!]
\centering\includegraphics{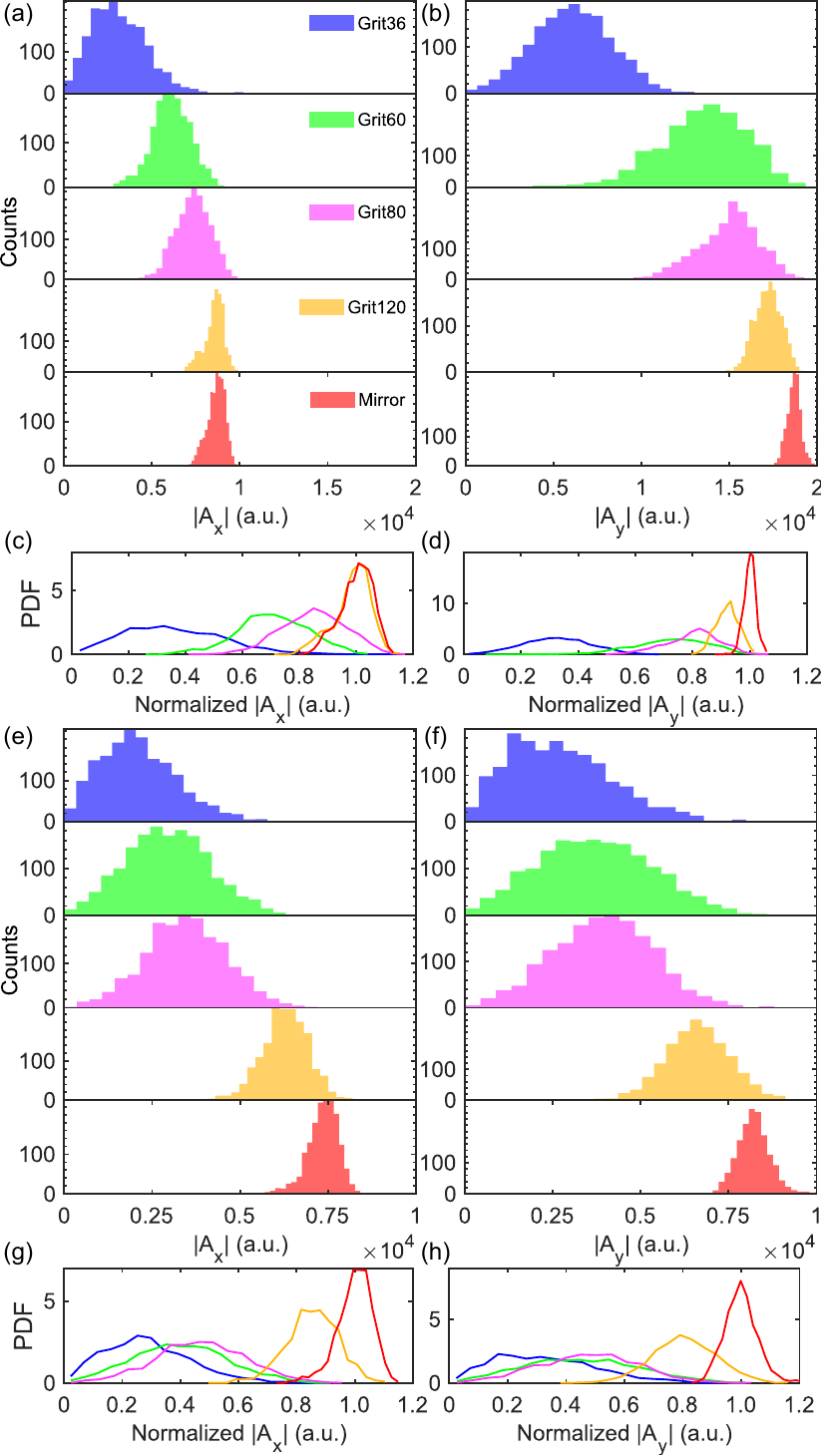}
\caption{Distribution of electric field amplitude along 
 the X and Y channels, $|\mathbf{A}_x|$ and $|\mathbf{A}_y|$, are shown in (a) and (b) respectively, at 0.3 THz using histograms for each gold-coated sandpaper and the reference mirror. The PDF of normalized $|\mathbf{A}_x|$ and $|\mathbf{A}_y|$ are summarized in (c) and (d) at 0.3 THz. The same histograms obtained for 0.5 THz are shown in (e) and (f), and the corresponding PDF in (g) and (h), respectively.}
\label{fig:pdf}
\end{figure}

\subsection{First-order Statistics of the THz Speckle Fields}
We investigated the quantitative relationship between the THz speckle images and the surface profile of the targets and the frequency of THz illumination. Since the images of the Stokes parameters are essentially the joint functions of two separate data sets, comprised of the complex Fourier coefficients of the orthogonal components of the THz electric fields, i.e., $\{\mathbf{A}_x\}$ and $\{\mathbf{A}_y\}$ (Eq. \ref{eq:1}), we start with the independent analysis of each, which is also referred to as the first-order statistics. For a fixed THz frequency, $\{\mathbf{A}_x\}$ or $\{\mathbf{A}_y\}$ would always consist of $N=1681$ complex scalars corresponding to different pixels in space. Figure \ref{fig:pdf} presents the distribution of $\{|\mathbf{A}_x|\}$ (left column) and $\{|\mathbf{A}_y|\}$ (right column) at 0.3 and 0.5 THz for the five targets. It is evident that with increasing surface roughness, these distributions vary from a narrow shape resembling a delta function to a Gaussian-like form of smaller mean value and larger standard deviation, and finally approaching the Rayleigh distribution. The absolute values of amplitudes in different histograms are usually not directly comparable due to the variation in the incident and backscattered THz intensity. Therefore, we have also calculated the PDF of normalized amplitudes in Figure \ref{fig:pdf}, showing that $\{\mathbf{A}_x\}$ and $\{\mathbf{A}_y\}$ follow a similar trend, i.e., the departure from the delta-like function increases with increasing roughness and frequency.

\begin{figure}[b!]
\centering\includegraphics{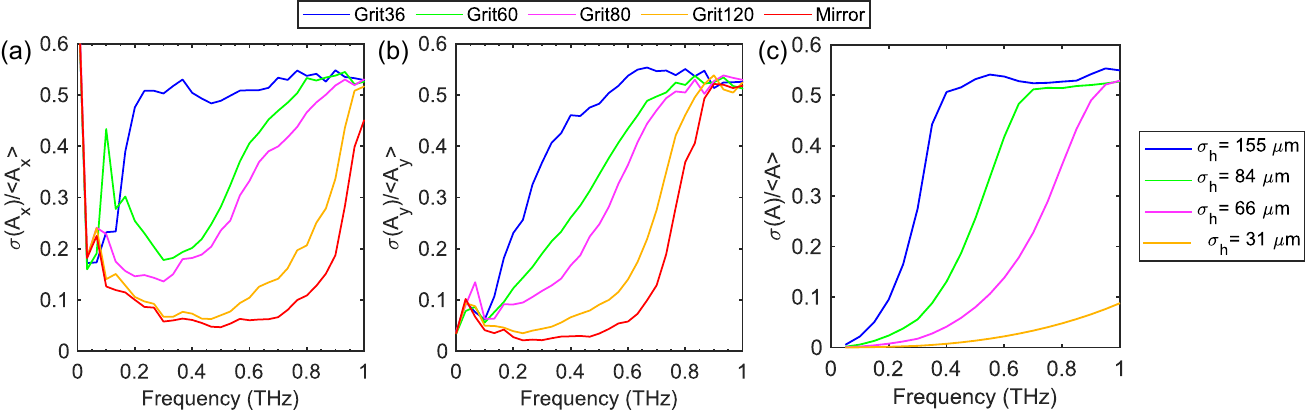}
\caption{Amplitude contrast of the speckles formed by different samples separated into two orthogonal polarizations of the THz waves, along the laboratory $x-$ (a) and $y-$ (b) directions. (c) Numerical simulation of the amplitude contrast of the speckles corresponding to different grits of sandpapers. The RMS height $\sigma_h$ used for simulation is obtained from \cite{sandpaper,sandpaper2}.}
\label{fig:contrast}
\end{figure}

The contrast of the THz amplitude, defined as the relative standard deviation (RSD) \cite{Goodman:20} of the probability distributions function, is presented in Fig. \ref{fig:contrast} in the frequency range between 0 and 1 THz. The RSD values are calculated by the ratio of the standard deviation of the THz amplitude to the mean of the same quantity. The measurements obtained from the reference mirror have non-zero contrast values varying with frequencies, which are attributed to the systematic errors of our instrument, such as the frequency-dependent performance of the PCAs, cross-talk between the channels and the reflections from the internal walls, the fluctuations of THz illumination with time, the polarization changes induced by our beam steering system \cite{Anzolin:10}, etc. The mirror contrast measurements shown by the red traces in Fig. \ref{fig:contrast}(a and b) can be used to determine a useful frequency range below 0.6 THz, in which RSD is still a good measure of the speckle noise relative to the surface roughness. For both X and Y detection channels, the amplitude contrast rise with the THz frequency and finally saturates around $\sqrt{(4-\pi)/\pi}\approx0.523$, which is the theoretical RSD value of a Rayleigh distribution. Also, the curves for targets with increased roughness have larger contrast and will reach the plateau value of 0.523 earlier, as compared to the smoother samples. In summary, upon the illumination from 0 to 0.6 THz, the gold-coated sandpaper sample of grit 120 only generated somewhat small amount of speckle, whereas the grit 80 and 60 samples have generated partially-developed speckles, and the grit 36 sample gives rise to fully-developed speckles at higher frequencies.

The contrast variations in Fig. \ref{fig:contrast} can be explained by 
the sums of finite random phasors with a non-uniform distribution of phase functions, which is a classic treatment for partially-developed speckles \cite{Pedersen:74,Goodman:75,Asakura:78}. We have adopted a numerical approach \cite{Bergoend:08}, based on generating a 2D height profiles with Gaussian correlation functions, and summing up the reflected wavefronts that has been dephased by the surface. The simulation results are summarized in Fig. \ref{fig:contrast}(c), showing great agreement with trends observed in Fig. \ref{fig:contrast}(a and b).

\subsection{Second-order Statistics of the THz Speckle Fields}

\begin{figure}[b!]
\centering\includegraphics{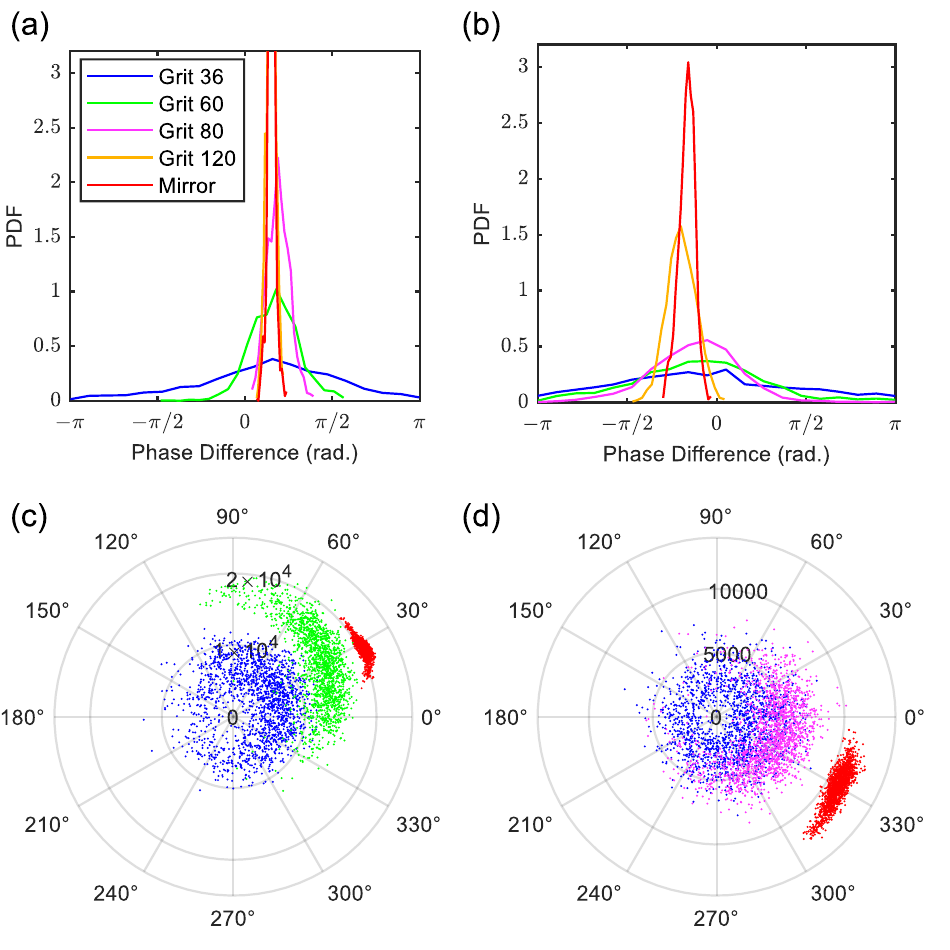}
\caption{Distribution of phase difference $\theta_{xy}$ for the speckles fields scattered by the five targets at (a) 0.3 THz and (b) 0.5 THz. Representation of speckles fields in the polar coordinates of $\sqrt{I}$ and $\theta_{xy}$ at (c) 0.3 THz and (d) 0.5 THz.}
\label{fig:phase}
\end{figure}

In addition to their modulus, the argument (or phase) of complex $\{\mathbf{A}_x\}$ or $\{\mathbf{A}_y\}$ is another important parameter for the statistical analysis of speckle patterns. However, the phase measurements obtained by a single channel THz-TDS system often suffer from errors caused by the time-shift of laser pulses (drift or jitter) or the placement of sample. Instead, we choose to investigate the difference of phase between $\{\mathbf{A}_x\}$ and $\{\mathbf{A}_y\}$, which reduces the uncontrolled phase shift occurred at the same point in time or space. Figure \ref{fig:phase}(a) and (b) show the PDF of $\theta_{xy}=\text{arg}(\mathbf{A}^*_y\mathbf{A}_x)$ for the five targets at 0.3 and 0.5 THz, respectively. 
For different THz frequencies, the PDF of ${\theta_{xy}}$ generally have quasi symmetric forms, while they become more dispersed over $[-\pi, \pi]$ as the roughness or frequency increases, and finally approach the uniform distribution. 

For the perturbative surface (Grit 120) ${\theta_{xy}}$ still center around similar value to that of incident THz beam (Mirror), yet for the rougher surface (e.g., Grit 36) the central value of ${\theta_{xy}}$ tend to approach zero. In simplified models, such as Kirchhoff Approximation \cite{Beckman:87} or Small Perturbation Method \cite{Valenzuela:67}, the normal-incidence backscatter from perfectly conducting, but slightly rough, surface is predicted to retain the incident polarization. Therefore, the trend between $\theta_{xy}$ and relative roughness shown in Fig. \ref{fig:phase} involves a scattering phenomena of higher complexity. Also, we have generated representations of speckle fields in the polar coordinates using $\sqrt{I}$ and $\theta_{xy}$, which are shown in Fig. \ref{fig:phase}(c) and (d). Taking advantage of the second-order statistics, these representations have provided adequate discrimination between different targets.

\begin{figure}[h]
\centering\includegraphics{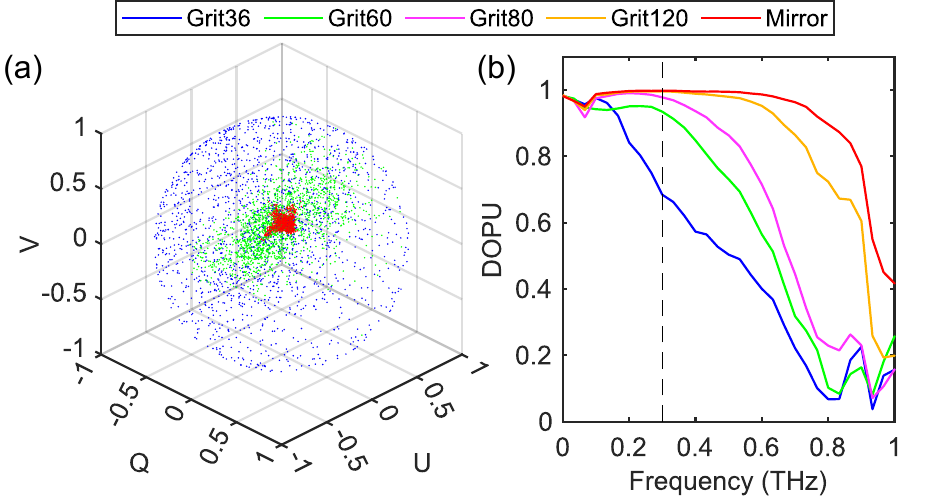}
\caption{(a) The distribution of normalized Stokes vectors on the Poincaré sphere for the speckles fields at 0.3 THz. (b) Frequency dependent degree of polarization uniformity (DOPU) for different samples.}
\label{fig:clouds}
\end{figure}

As defined in Eq. \ref{eq:1}, the Stokes parameters also contain the statistics of $\theta_{xy}$, since $\theta_{xy}=\tan^{-1}(V/U)$. Figure \ref{fig:clouds}(a) show the Poincaré sphere representation of the normalized Stokes vectors of the speckles fields scattered by sandpapers of grit 36, grit 60 and flat mirror at 0.3 THz. It is clear that the the Stokes vectors become more dispersed on the unit sphere as the  target roughness increases, which simultaneously decreases the norm of the average Stokes vectors from 1 to 0. This norm is thus a good measure of spatial randomness of the polarization speckles, which has been named by Wang et al. \cite{Wang:08} as the "spatial degree of polarization" and by Götzinger et al. \cite{Gotzinger:08} as the "degree of polarization uniformity (DOPU)". For simplicity, we will use the abbreviation DOPU defined as,
\begin{equation}
\centering
\text{DOPU}=\sqrt{\overline{Q}^2+\overline{U}^2+\overline{V}^2}/\overline{I},
\label{eq:3}
\end{equation}
where the spatial average operation is applied over the N = 1681 pixels in the entire 20×20 $mm^2$ images. There is a slightly different definition of DOPU than given in Eq. \ref{eq:3}, where the normalization by $I$ is applied before the spatial averaging operation \cite{Gotzinger:08}, yet we did not find any significant change using our data sets. Figure \ref{fig:clouds}(b) presents the frequency dependent DOPU values for the five targets. 
Despite the observable variations in the Stokes parameters of the reference mirror (as shown in Fig. \ref{fig:IQUV}), we determined that $\text{DOPU}\in[0.98, 1]$ is valid in the frequency range between 0.1 and 0.6 THz, corresponding to the high uniformity of fully-polarized fields. Instead, the departure of DOPU from 1 at higher frequencies suggests the limitation of our instrument due to other sources of polarization measurement noise, as explained earlier, rather than the speckles. As for the gold-coated sandpapers, DOPU gradually approaches 0 with increased roughness and THz frequency, indicating that the randomness of THz polarization speckles are associate with the strength of scattering.

The statistics of $\theta_{xy}$ can be attributed to the scattering matrices $\mathbf{S}$, expressed by
\begin{equation}
\centering
\begin{bmatrix}
\mathbf{A}_x\\
\mathbf{A}_y
\end{bmatrix}=\begin{bmatrix}
\mathbf{S}_{xx} & \mathbf{S}_{xy}\\
\mathbf{S}_{yx} & \mathbf{S}_{yy}
\end{bmatrix}\begin{bmatrix}
\mathbf{A}_{x}^{i}\\
\mathbf{A}_{y}^{i}
\end{bmatrix},
\label{eq:2}
\end{equation}
where $i$ labels the quantities for incident waves. 
For a smooth surface, $\mathbf{S}$ is the identity matrix and no change in $\theta_{xy}$ is expected. As for the rough surface, It has been substantiated that two types of changes in $\mathbf{S}$ can both cause $\theta_{xy}$ to vary: (i) the cross-polarization entries $\mathbf{S}_{xy}$ and $\mathbf{S}_{yx}$ become non-zero \cite{Vesperinas:82,Liu:15}. (ii) the co-polarization entries $\mathbf{S}_{xx}$ and $\mathbf{S}_{yy}$ become out of phase (i.e. $\text{arg}(\mathbf{S}_{xx}-\mathbf{S}_{yy})\neq0$) \cite{ODonnell:91,Touzi:96}. Therefore, characterizing the complete matrix $\mathbf{S}$ is necessary for understanding which mechanism is underlying in our experimental conditions. 
Experimental conditions that can distinguish between these two mechanisms would thus require a higher degree of the polarization control of the THz emission and calibration of the systematic errors induced by our instrument, which represent the limitation of our current PHASR Scanner design.

\section{conclusion}
We have presented a handheld and fast polarization-sensitive THz spectral imaging method for resolving speckle fields in terms of the Stokes vectors of the backscattered light. This method requires two PCAs and a polarizing beam-splitter to be incorporated into the PHASR Scanner we have previously developed. We statistically analyzed the Stokes vector images formed by the new instrument from the gold-coated sandpapers of different grit sizes. The first-order statistics of the THz speckle fields gradually transition to Rayleigh distribution as the target roughness becomes comparable to the wavelength of light, which is predicted by the models of partially-developed speckles. The second-order statistics of the THz speckle fields, i.e., the polarization states, reveal that the randomness of phase difference or Stokes vectors serve as an accurate measure of the strength of scattering. This work can pave the way for THz polarimetric imaging of speckle patterns as a potential marker for discrimination of sample-induced scattering in biomedical imaging and non-destructive testing.

\begin{backmatter}
\bmsection{Funding} Stony Brook University; National Institute of General Medical Sciences (GM112693).


\bmsection{Disclosures} The authors declare no conflicts of interest.

\bmsection{Data availability} Data underlying the results presented in this paper are not publicly available at this time but may be obtained from the authors upon reasonable request.


\end{backmatter}

\bibliography{citation}

\end{document}